\documentclass[10pt,english]{article}
\usepackage{subfigure}
\usepackage{graphicx}
\usepackage{float}
\usepackage[T1]{fontenc}
\usepackage[latin9]{inputenc}
\usepackage[margin=1.25in]{geometry}
\usepackage{float}
\usepackage{amsmath}
\usepackage{amsbsy}
\usepackage{setspace}
\usepackage{amssymb}
\usepackage{esint}
\usepackage{cite}
\usepackage{hyperref}

\usepackage{listings}
\usepackage{xcolor}
\definecolor{codegreen}{rgb}{0,0.6,0}
\definecolor{codegray}{rgb}{0.5,0.5,0.5}
\definecolor{codepurple}{rgb}{0.58,0,0.82}
\definecolor{backcolour}{rgb}{0.95,0.95,0.92}

\lstdefinestyle{mystyle}{
    backgroundcolor=\color{backcolour},   
    commentstyle=\color{codegreen},
    keywordstyle=\color{magenta},
    numberstyle=\tiny\color{codegray},
    stringstyle=\color{codepurple},
    basicstyle=\ttfamily\footnotesize,
    breakatwhitespace=false,         
    breaklines=true,                 
    captionpos=b,                    
    keepspaces=true,                 
    numbersep=5pt,                  
    showspaces=false,                
    showstringspaces=false,
    showtabs=false,                  
    tabsize=2
}
\makeatletter

\begin{document}

\title{Raspberry Pi Pico as a Radio Transmitter}

\author{M. Andrecut}

\date{August 22, 2025}

\maketitle
{

\centering Unlimited Analytics Inc.

\centering Calgary, Alberta, Canada

\centering mircea.andrecut@gmail.com

} 

\bigskip

\bigskip

\bigskip

\begin{abstract} 
In this paper we discuss several surprisingly simple methods for transforming the Raspberry Pi Pico (RP2) microcontroller into a radio transmitter, 
by using only cheap off the shelf electronic components, and open source software. 
While initially this transformation may look as a harmless curiosity, in some extreme cases it can also pose security risks, since it can be used to open a large number of local stealth radio communication channels.  
\end{abstract}

\bigskip

\bigskip

\section{Introduction}

The Raspberry Pi Pico (RP2) is a very popular and cheap microcontroller frequently used in IOT (Internet of Things) applications \cite{key-1}.  
While the RP2 microcontroller comes in several flavors, such as the more expensive version W, containing the additional Wi-Fi and Bluetooth connectivity modules \cite{key-2}, here we limit ours investigation to the 
cheapest and most stripped down version. This cheap RP2 version supposedly lacks any wireless communication capabilities, 
however here we will show how it can be easily transformed into a radio transmitter, just by 
using very cheap, off the shelf electronic components and already existing free open source software. 
This transformation initially may look as a harmless curiosity, however in some extreme cases it can also pose security risks, since it can be used to open a large number of local stealth radio communication channels.

While the project sounds exciting, and one can be very eager to try it out immediately, we should warn the reader that strong radio transmissions require a permit. 
Also, the legislation varies from country to country, and to avoid any potential problems it is recommended to minimize the transmission time and range to several seconds and meters, 
and to use the described techniques only for experimental purposes, avoiding as much as possible to cause any significant disruption. 
The scope of this work is research and educational, and the author is not responsible for the non compliant use of the techniques described here. 

\newpage

\section{The RP2}

The RP2 is a low-cost, high-performance microcontroller board using the ARM Cortex-M0+ RP2040 chip produced by the Raspberry Pi Foundation \cite{key-1}, \cite{key-2}. 
The RP2040 chip is a very efficient 133MHz dual-core CPU, and incorporates 264KB of SRAM, 2MB of Flash memory, accurate timing modules, 26 GPIO (General-Input-Output) pins, 3 ADC pins (Analog-to-Digital Converters), 
8xPIO (Programmable I/O) state machines for custom peripherals, and commonly used peripheral interface modules such as: 2xUART (Universal Asynchronous Receiver-Transmitter), 16xPWM (Pulse Width Modulation) channels, 
2xSPI (Serial Peripheral Interface), 2xI2C (Inter-Integrated Circuit). 
However, in its most basic version which we will use here, the board does not offer any radio communication capabilities, such as Wi-Fi or Bluetooth. 

The RP2 can be programmed in C/C++ or MicroPython. In this paper we use MicroPython \cite{key-3}, since it is more accessible. 
MicroPython's module provides low level functions related to the hardware on the microcontroller board \cite{key-4}. 
Some of these functions have direct and unrestricted hardware access to the CPU, timers, buses etc. An incorrect use may
lead to malfunction, lockups, crashes, and hardware damage in some extreme cases. 
From the MicroPython point of view the RP2 board can be controlled using eleven different software classes. 
Here we will only use the following classes: 
\begin{itemize}
\item Pin - control of the I/O pins.

A Pin object is used to control GPIO, and it has methods to set the mode of the pin as IN or OUT, and to get and set the digital logic level. 
The RP2W has 26 multifunction GPIO pins connectable to external devices.

\item PWM - pulse width modulation.

The signal sent to an output pin is either HIGH or LOW voltage. PWM it is
used to control the amount of time a signal is HIGH. 
There are 8 independent PWM generators called slices, each having two channels. 
The generators can be clocked from 8Hz to 62.5Mhz, at a machine frequency of
125Mhz. While the two channels of a slice will run at the same frequency, they can
have a different duty rate. The two channels are usually assigned to adjacent GPIO
pin pairs with even/odd numbers: slice 0 (GPIO0, GPIO1), slice 1 (GPIO2, GPIO3), and so on.

\item ADC - analog to digital conversion.

We can use this class to sample continuous voltages and convert them to discrete values. The standard ADC input
value range is 0-3.3V. There are four ADC channels, where the input signal for ADC0, ADC1, ADC2 and ADC3 can be connected with: GPIO26, GPIO27, GPIO28, and GPIO29.

\end{itemize}

\section{Radio configuration test}

The \textit{metamorphosis} of the stripped down  RP2 into a radio transmitter is surprisingly simple, and it only requires the use of PWM to generate both the modulator and the carrier signals, 
a capacitor to connect the signals, and a wire as an antenna. 

For experimental purposes we would like to limit our transmission range to a small local radius (around 10-20m), and therefore here we will use a simple antenna consisting of a short wire of about 10cm. 
The basic connectivity schema is shown in Figure 1. The RP2 is powered by using the USB connector, and this ensures that the VBUS is at Vcc=5V, and all GND pins are connected to the ground. 
Here we simply connect the GPIO0 and GPIO2 using a capacitor $\text{C}\simeq 1 - 100\mu\text{F}$, and we attach the short wire antenna to the GPIO0. 
Also, we can use the Thonny IDE \cite{key-5} to write a simple Python test program, and run it on RP2 (Listing 1). 
\begin{figure}[!ht]
\centering \includegraphics[width=7cm]{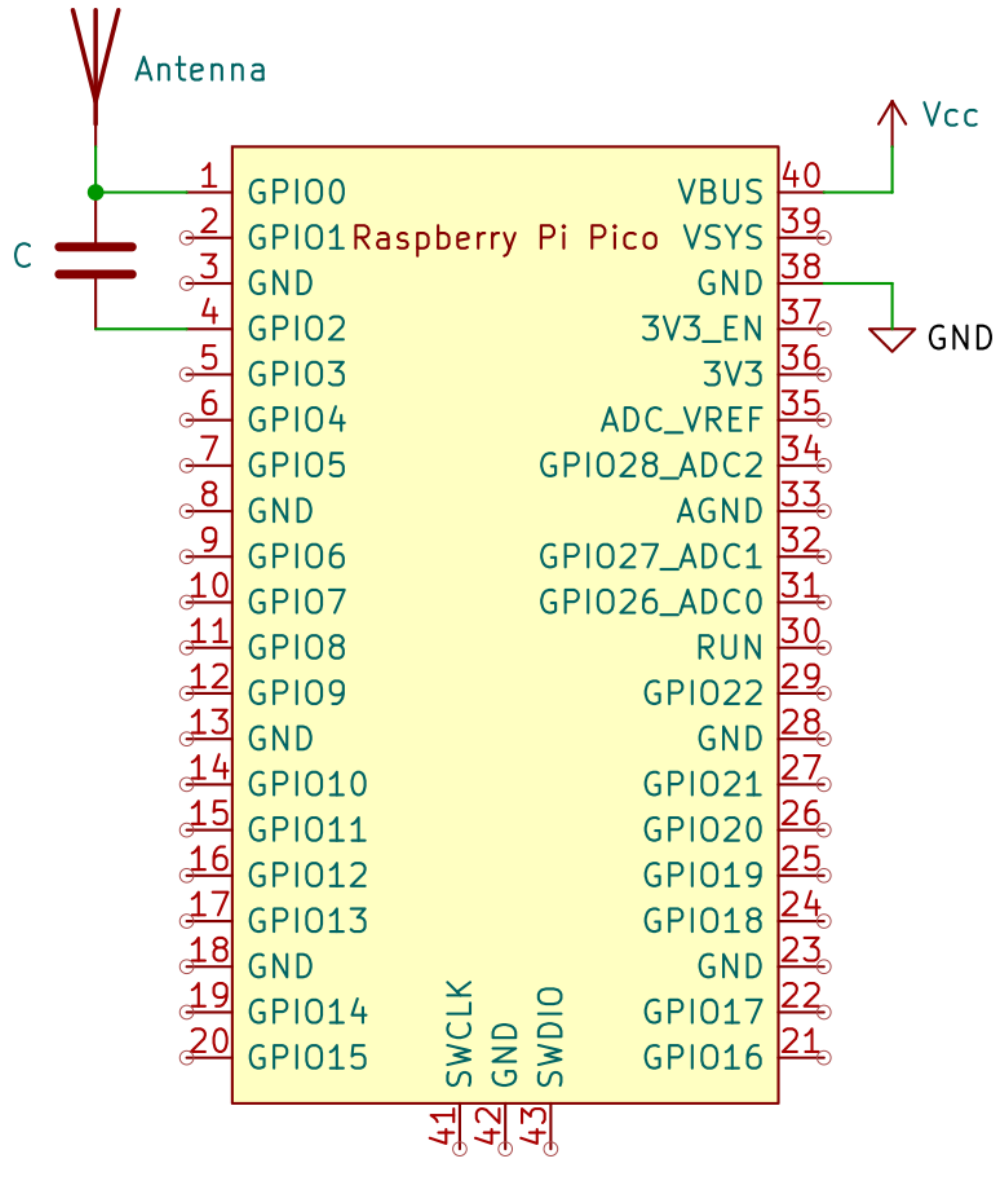}
\caption{RP2 configuration.}
\end{figure}

\begin{lstlisting}[language=Python,caption=Initial RP2 test]
import time
from machine import Pin, PWM

pwm0 = PWM(Pin(0))
pwm0.duty_u16(16384)
pwm0.freq(31250000)
pwm1 = PWM(Pin(2))
pwm1.duty_u16(32768)
while True:
    for n in range(100):
        pwm1.freq(10*n + 200)
        time.sleep(0.01)
\end{lstlisting}

The program generates a carrier signal with the frequency $\nu^*=31.25$MHz on GPIO0, and a linear frequency sweep signal with a frequency $\nu = (10n+200)$Hz, $n=0,1,...,99$, and a duration of 1s on GPIO2. 
The carrier signal is then modulated by the sweep signal on GPIO2 via the capacitor connection C. The sweep signal is repeated until the program is stopped from the Thonny IDE. 

Now that we have the sweep signal generated and radio transmitted, we should be able to tune the radio on 31.25MHz and hear it. Unfortunately, the FM radio band range is between 88MHz to 108MHz, so unless 
you have a radio that can be tuned outside of the FM band you won't be able to hear anything. 

We should note that while this method of using rectangular PWM signals for both the carrier and the modulation does generate a radio transmission, it also creates  harmonics in the radio spectrum. 
It is well known that a rectangular wave signal is made up of an infinite number of sine waves, starting at the fundamental frequency and then going up with all the odd numbered harmonics \cite{key-4}. 
In general, a rectangular PWM signal with the period $T=2L=1/\nu$ can be defined as following:
\begin{equation}
\Pi(t) = 2[H(t/L) - H(t/L-1)] - 1,
\end{equation}
where $H(x)$ is the Heaviside step function:
\begin{equation}
H(x) = 
  \begin{cases}
   0  & \text{for } x < 0 \\
   1/2 & \text{for } x=0 \\
   1 & \text{for } x > 0
  \end{cases}
\end{equation}

The Fourier series of a periodic signal $f(t)$ is defined as an infinite sum of sines and cosines:
\begin{equation}
f(t) = a_0 + \sum_{n=1}^\infty [a_n \cos(n t) + b_n \sin (n t)].
\end{equation}
For a rectangular function we have $a_0=a_n=0$, since the rectangular function is odd. 
Therefore only the $b_n$ coefficients need to be calculated, and they are given by: 
\begin{equation}
b_n = \frac{1}{L} \int_0^{2L} \Pi(t) \sin \left( \frac{n\pi t}{L} \right) dt = \frac{4}{n\pi} \sin^2 \left( \frac {n\pi}{2} \right) 
= 
  \begin{cases}
   0  & \text{for } n = 2k \\
   \frac{4}{n\pi} & \text{for } n = 2k+1
  \end{cases}
\end{equation}
such that we have:
\begin{equation}
\Pi(t) = \frac{4}{\pi} \sum_{k=0}^\infty \frac{1}{2k+1} \sin \left( \frac{(2k+1)\pi t}{L} \right).
\end{equation}
In principle, the unwanted harmonics in the transmission can be attenuated using band filters \cite{key-4}. However, to keep things simple, and also as disruptive as possible, here we continue without the band filters.

So, how can we hear the radio transmission if the frequency falls outside of the standard radio bands? 
The simplest way to approach this is by using an RTL-SDR receiver, which can be connected to a computer or to a cellular phone using an USB connector.  

The RTL-SDR receiver is built around the RTL2832U and R820T chips, which are a high-performance demodulator and tuner, initialy developed for the digital TV industry. 
The RTL-SDR dongles can receive radio frequencies from 500kHz up to 1.75GHz, depending on the particular model, and here we will use again the cheapest version.

The RTL-SDR receiver can be used as a computer based radio receiver, with a software defined radio (SDR) controller. 
The software for the RTL-SDR is community developed and open source, and it is freely available on all the common computer platforms \cite{key-6}. 
Also, one can install the RTL-SDR software on a cell phone, tablet, or a laptop, for creating a mobile radio receiver. 

\begin{figure}[!ht]
\centering \includegraphics[width=15.25cm]{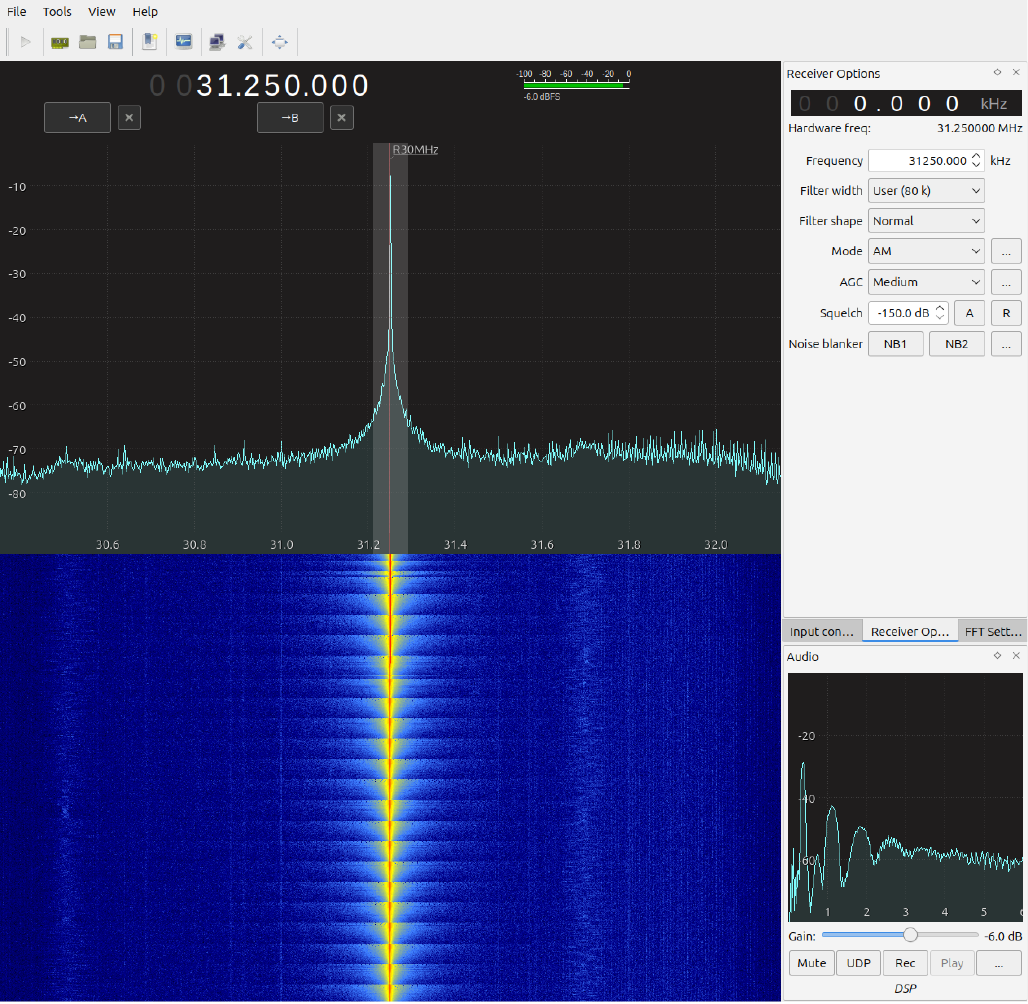}
\caption{RTL-SDR receiver, sweep signal test.}
\end{figure}

\bigskip

Here we will use the popular GQRX receiver \cite{key-7} which is powered by the GNU radio \cite{key-8}, but one can use any other SDR receiver available, and obtain similar results. 
Once the software is properly installed, and the RTL-SDR receiver recognized, one can tune on the carier frequency, or on any of the harmonic frequencies that fall in the bandwidth of RTL-SDR. 
For example, in my area the signal is relatively clean and strong at the carrier frequency 31.25MHz. 

In the Receiver Options of the GQRX, we should select the following options: 
Frequency: 31250.000KHz, Filter width: User (80k) (drag the window to make it as large as possible), Filter shape: Normal, Mode: AM, AGC: Medium. Also, in the Input Controls tab set the LNA around 24.8 db, No limits, and DC remove. 
The rest of the controls can remain as they are. 
You will see a quite strong signal in the FFT (Fast Fourier Transform) window (Figure 2). Even more surprisingly, if you turn on the volume you will hear the sound made by the sweep signal, which resembles the sound 
of an alarm. 
Next, try to tune the frequency to multiples of 31.25MHz, and you will notice that you can see and hear the same signal. 
Now that we have a successful proof of concept we can experiment with some more complex and interesting transmissions. 

\section{Morse code radio transmitter}

Using the previously described PWM modulation method we can also transmit Morse code \cite{key-9}, which could be useful for stealth communications. 
Let us first discuss some details about the Morse code, for the readers who are not familiar with it. 

Morse Code is one of the oldest radio communication methods. Basically, the Morse code is a text encoding and decoding method, which can be used in radio transmissions. 
Using the Morse code, the letters, the numbers and the punctuation characters, can be encoded in units of transmission time dt as following: 
1 (dit='.'); 3 (dah='-'); 1 (spacing between dit and dah); 3 (spacing between the characters of a word); 7 (spacing between words). 
For example the word PARIS consists of 50 units of time, encoded as following:

\bigskip

P='.--.', 1131311, 3 (14 time units)

A='.-', 113, 3 (8 time units)

R='.-.', 11311, 3 (10 time units)

I='..', 111, 3 (6 time units)

S='...', 11111, 7 (12 time units)

\bigskip

\noindent One can see that at the end of a word character we insert an inter-character time (3), and at the end of a word we insert an inter-word time (7).

The Morse encoding dictionary (MorseCode) is included in the example Listing 2, where we use MicroPython to transmit the "Hello World!" string (or any other text message) with the carrier frequency of 31.25MHz. 
Also, as mentioned before, an important parameter is the unit of transmission time, which in this example was set to dt=0.05s, but it can be changed to make the transmission faster or slower. 

First the string "Hello World!" is converted to upper case (Morse code only uses upper case letters), then it is encoded 
into the Morse code using the Morse dictionary and the corresponding time units using the function "morse\_encode()". The resulted Morse encoded text is then 
radio transmitted using the "morse\_broadcast()" function, with the same PWM carrier signal of 31.25MHz on GPIO0, modulated with a $600$Hz PWM signal on GPIO2 (duty=$50\%$), 
using the same circuit from Figure 1. 

\begin{lstlisting}[language=Python,caption=Morse code transmitter.]
from machine import Pin,PWM
import time

MorseCode = {
    'A': '.-', 'B': '-...', 'C': '-.-.', 'D': '-..', 
    'E': '.', 'F': '..-.', 'G': '--.', 'H': '....', 
    'I': '..', 'J': '.---', 'K': '-.-', 'L': '.-..', 
    'M': '--', 'N': '-.', 'O': '---', 'P': '.--.', 
    'Q': '--.-', 'R': '.-.', 'S': '...', 'T': '-',
    'U': '..-', 'V': '...-', 'W': '.--', 'X': '-..-', 
    'Y': '-.--', 'Z': '--..',
    '0': '-----', '1': '.----', '2': '..---', 
    '3': '...--', '4': '....-', '5': '.....', 
    '6': '-....', '7': '--...', '8': '---..', 
    '9': '----.',
    '.': '.-.-.-', ',': '--..--', '?': '..--..', 
    "'": '.----.', '!': '-.-.--', '/': '-..-.', 
    '(': '-.--.', ')': '-.--.-', '&': '.-...', 
    ':': '---...', ';': '-.-.-.', '=': '-...-',
    '+': '.-.-.', '-': '-....-', '_': '..--.-', 
    '"': '.-..-.', '$': '...-..-', '@': '.--.-.', 
    ' ': '/',  
}

def morse_encode(text):
    x,dt = [],0.05
    for ch in text:
        for c in MorseCode[ch]:
            if c == '-':
                x.append((1,3*dt))
                x.append((0,dt))            
            elif c == '.':
                x.append((1,dt))
                x.append((0,dt))            
            elif c == '/':
                x.append((0,6*dt))                    
        x.append((0,2*dt))                    
    return x

def morse_broadcast(encoded_text):    
    led = Pin(25, Pin.OUT) # Pico
#    led = Pin("LED", Pin.OUT) # Pico W
    for x in encoded_text:
        if x[0] == 1:
            led.on()
            pwm1.duty_u16(32767)
        else:
            led.off()
            pwm1.duty_u16(0)
        time.sleep(x[1])

if __name__ == "__main__":
        
    text = 'Hello World!'
    text = text.upper()
    print(text)    

    encoded_text = morse_encode(text)
    print(encoded_text)
    
    pwm0 = PWM(Pin(0))
    pwm0.freq(31250000)
    pwm0.duty_u16(32768)

    pwm1 = PWM(Pin(2))
    pwm1.duty_u16(32768)
    pwm1.freq(600)    

    while True:
        morse_broadcast(encoded_text)
\end{lstlisting}

The next question is: how can we receive and decode the Morse code message? 
This is a bit more complex problem than it looks, and besides the previously mentioned GQRX, it requires several other software packages, also open source and freely available on any software platform, such as: 
\begin{itemize}
\item nc \cite{key-10} - also known as Netcat, is a command-line utility used for reading from and writing to network connections using TCP or UDP (here we will use UDP). 
\item sox \cite{key-11} - is a command-line audio processing tool, particularly suited for batch processing. 
\item multimon-ng \cite{key-12} - can decode a variety of digital radio transmission modes.
\end{itemize}

First we need to start GQRX using the same settings (31.25MHz) as before, with the exception of the Filter width, which now should be set to Normal or Narrow. 
In the Receiver Options select UDP for the communication protocol. Then GQRX will use the port 7355 on the computer to pipe the received audio signal. 
This signal is then processed using the following command (Listing 3).
\begin{lstlisting}[language=Bash,caption=Morse code decoding.]
nc -l -u 7355 | sox -r 48000 -t raw -b 16 -c 1 -e signed-integer /dev/stdin -r 22000 -t raw -b 16 -c 1 -e signed-integer - | multimon-ng -t raw -a MORSE_CW /dev/stdin
\end{lstlisting}
and "voila", we can decode the Morse code message radio transmitted by the RP2 (Listing 4).
\begin{lstlisting}[language=Bash,caption=Morse code decoded message.]
multimon-ng 1.3.0
...
Enabled demodulators: MORSE_CW
HELLO WORLD!HELLO WORLD!HELLO WORLD!HELLO WORLD!HELLO WORLD!HELLO WORLD!HELL 
\end{lstlisting}
Let us see what the command actually does: (1) nc -l -u 7355,  listens for UDP traffic on port 7355 configured in GQRX; 
(2) sox -r 48000 -t raw -b 16 -c 1 -e signed-integer /dev/stdin -r 22000 -t raw -b 16 -c 1 -e signed-integer -, 
resamples the audio received from netcat (via the pipe operator) to a sample rate suitable for multimon-ng (e.g., 22000) and ensures the correct audio format; 
(3) multimon-ng -t raw -a MORSE\_CW /dev/stdin, reads raw audio from standard input (/dev/stdin) and activates the MORSE\_CW decoder.

Because of the rectangle PWM signals used, the Morse message is disruptive, since it is also transmitted at multiples of the 31.25MHz frequency.
If necessary, the Morse code can be also easily obfuscated (encrypted), for example by randomly shuffling the dictionary. This way, only the receivers that also have the obfuscated dictionary 
will be able to decode the message. 

\section{Sequencer music broadcasting}

We can use the same basic configuration of the RP2 to broadcast sequencer music. To illustrate this approach, here we will use the "buzer\_music" package \cite{key-13}, which is a 
MicroPython library to play music through a buzzer, and replaces chords with fast arpegios to simulate polyphony. 
Contrary to its initial purpose, we will not use the library to play music through a buzzer, but we will use it to modulate the PWM carrier signal, 
such that we can broadcast the music on the 31.25MHz frequency (and its multiples). The result could be a radio station playing favorite sequencer music files. 
The MicroPython code is given in the Listing 3. 

\begin{lstlisting}[language=Python,caption=Sequencer music broadcasting.]
from buzzer_music import music
from time import sleep
from machine import Pin, PWM

pwm0 = PWM(Pin(0))
pwm0.freq(31250000)
pwm0.duty_u16(16384)

song = open("seq_music.txt", "r").read()
mySong = music(song[25:-2], pins=[Pin(2)])

while True:
    print(mySong.tick())
    sleep(0.04)
\end{lstlisting}

The sequencer music can be download from \url{onlinesequencer.net}. For example here we will use the song: \url{(https://onlinesequencer.net/1696155} (Undertale - Heartache). 
On this webpage we have to click edit, select all notes with CTRL+A and then copy them with CTRL+C, paste the string in a simple text editor (notepad, gedit,...),  
and save it as a text file "seq\_music.txt", then copy the file to the RP2. 
We should notice that here we set the PWM carrier signal duty to 16384 ($25\%$). Also, we have removed the header (the first 25 characters) and the last 2 characters of the music string before the broadcasting. 
Next, tune GQRX on 31.25MHz and enjoy the music. 

\section{Recorded audio broadcasting}

In the previous section we have shown how to broadcast sequencer music, now we will extend this approach to broadcast recorded audio, for example "wav" files. 
To illustrate this approach, here we use the "mgm\_music.wav" file containing the well known piece of music running at the beginning of the 20th Century Fox and MGM movies \cite{key-14}. 
Unfortunately we cannot play directly the file on the RP2. 
However, here we will use a quite simple workaround, which only requires the  
conversion of the file into a "raw" file using FFmpeg (Listing 6), which is a well known open source cross-platform solution to record, convert and stream audio and video \cite{key-15}. 
\begin{lstlisting}[language=Bash,caption=FFmpeg conversion from "wav" to "raw".]
ffmpeg -i mgm_music.wav -ar 16000 -acodec pcm_u8 -f u8 mgm_music.raw
\end{lstlisting}
Here, "-i" specifies the input file "mgm\_music.wav", "-ar 16000" is the sample rate, "-acodec pcm\_u8" specifies the audio codec, "-f u8" specifies the 8bit conversion output to "mgm\_music.raw". 
After the "raw" conversion the music has an 8bit representation. 
We copy the "mgm\_music.raw" on the RP2, and we run the program from Listing 4. 

\begin{lstlisting}[language=Python,caption=Recorded audio broadcasting.]
import time
from machine import Pin, PWM

pwm0 = PWM(Pin(0))
pwm0.freq(31250000)
pwm0.duty_u16(16384)

pwm1 = PWM(Pin(2))
pwm1.freq(1000000)

buf = bytearray(4096)
while True:
    f = open("mgm_music.raw","rb")
    while f.readinto(buf) > 0:
        for sample in buf:
            pwm1.duty_u16(sample<<8)
            time.sleep_us(85)
    f.close()
\end{lstlisting}

We continuously read 4096 samples from the file, then each sample is multiplied with $2^8=256$, and we use the result to change the duty of the PWM modulator (pwm1). 
Here, "<<" represents the bitwise left shift operator, and in this case we perform an 8 bit left shift, which is equivalent to multiplying the original sample by $2^8=256$, but the shift operator is faster than the multiplication.
After each sample the program sleeps for $85\mu\text{s}$. This duration can be changed if you want to play the file slower or faster. In general, the optimal duration is a function of the sampling rate, so if we change the sampling rate 
we may have to change the sleep duration. It is surprising to hear how well the music sounds on the radio at 31.25MHz. 
Instead of music, one can also record, convert, and broadcast any kind of radio messages.

\section{Microphone radio bug}

Let us now see how we can transform RP2 into a microphone radio bug, or into a live radio transmitter. 
To accomplish this we will need a microphone, and here we will use the MAX4466 electret microphone module, which is also very cheap and widely available \cite{key-16}. Obviously the results will improve with a 
more expensive and better amplified product, but as a proof of concept we prefer to keep it cheap and simple. 
In Figure 3 we give the circuit schematic. The electret microphone MAX4466 is powered through the 3V3 pin, and grounded on AGND pin, which is the recommended ground for ADC. The output of the microphone is connected  
to the RP2 on the GPIO27\_ADC1. The MicroPython code is given in Listing 8. The rest of the radio transmitter is the same as before.

\begin{figure}[!ht]
\centering \includegraphics[width=9cm]{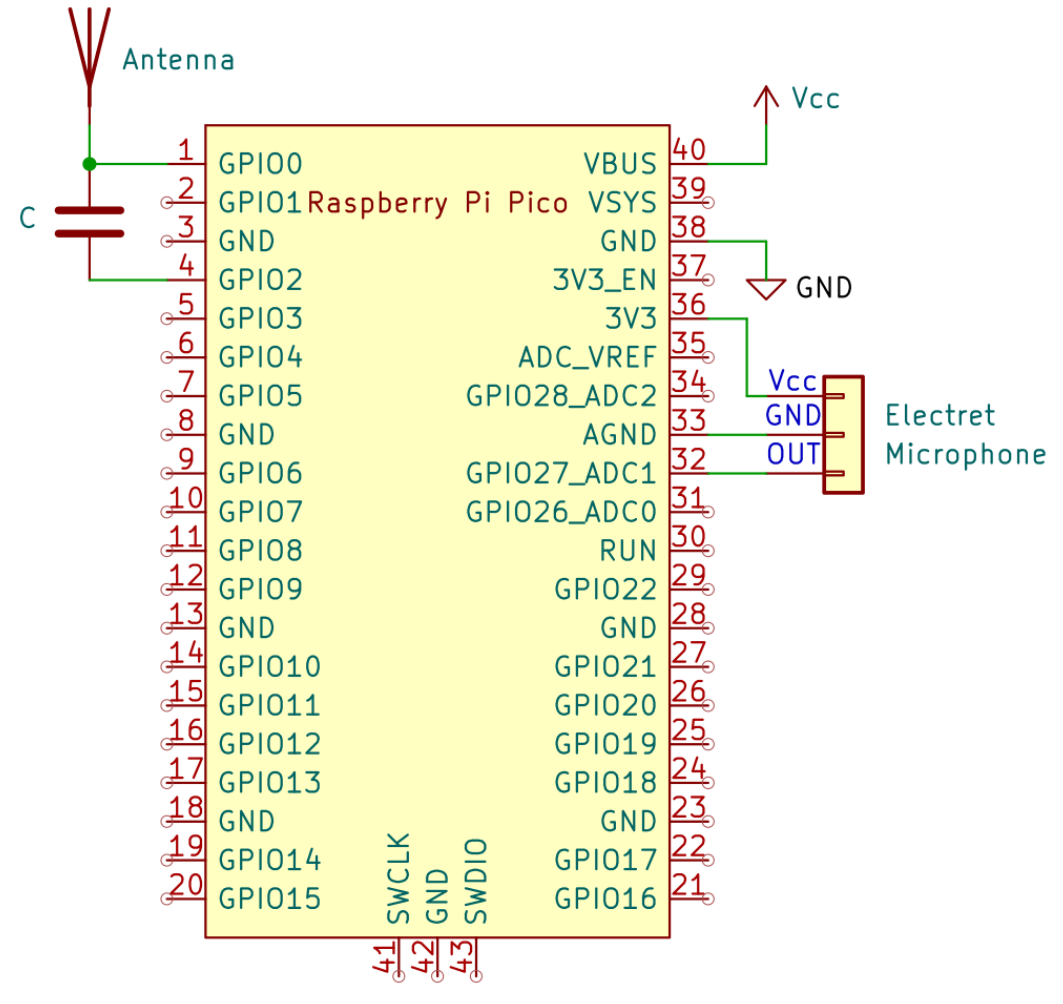}
\caption{Microphone radio bug.}
\end{figure}
 
\begin{lstlisting}[language=Python,caption=Microphone radio bug.]
from machine import Pin, ADC, PWM

pwm0 = PWM(Pin(0))
pwm0.freq(31250000)
pwm0.duty_u16(32768)

pwm = PWM(Pin(2))
pwm.freq(1000000)

adc = ADC(27)
while True:
    pwm.duty_u16(adc.read_u16())
\end{lstlisting}

The output of the MAX4466 electret microphone is digitally converted by the ADC using the GPIO27\_ADC1 connection, such that the resulted amplitude digital values are used to specify the duty value of the 
PWM modulator running on GPIO2. This signal is then used to modulate the PWM high frequency signal on GPIO0, by using the capacitor C. 
The microphone will pick up any local audio (such as conversations) and radio broadcast it. The result is a surprisingly clear and strong audio reception on 31.25MHz. 

\section{Conclusion}

In this paper we have shown how surprisingly simple is to transform the Raspberry Pi Pico (RP2) microcontroller into a radio transmitter, 
by simply using cheap off the shelf electronic components and free open source software. 
While in our approach we have limited the range of the transmission by using a short 10cm antenna wire, the range can be increased by simply increasing the length of the antenna. 
The optimal length of the antenna can be estimated with the formula $L=142.6/\nu^*$, where $\nu^*$ is the carrier frequency in MHz. 
For example, an optimal transmission on $\nu^*=31.25$MHz, requires a half-wave dipole antenna with a length of approximately 4.56m. 
Obviously, for some applications this length may be too much, so lengths of 1/4, 1/8, 1/16 the wavelength may be a good starting point. 
Proper antenna shielding and grounding can also be used to enhance the transmission performance, and reduce the interference. 
If a longer transmission range is required one can use an additional RF amplifier and a better antenna, which are also available at low cost. 

We should note that while these relatively simple transformation techniques have been illustrated using the RP2, they can be applied to any other microcontroller, or computer board 
that offers PWM output signals with frequencies in the RF range. 
Therefore, the relatively simple transformation of these devices into radio transmitters can pose security risks in some extreme cases, 
since they can be easily used to open a large number of local stealth radio communication channels. 

In closing, we should stress once again that it is always essential to comply with local regulations regarding radio frequency transmission. 

\section{Appendix}

The Github repository with the code and the data used in the paper is available at: 

\url{https://github.com/mandrecut/rp2_radio_transmitter}


\begin{thebibliography}{99}

\bibitem{key-1}
RP2040 Datasheet, Raspberry Pi Ltd. (2024).

\bibitem{key-2}
Raspberry Pi Pico W Datasheet, Raspberry Pi Ltd. (2024).

\bibitem{key-3}
Raspberry Pi Pico-series Python SDK, Raspberry Pi Ltd. (2024).

\bibitem{key-4}
M. Andrecut, \textit{Connecting with the Raspberry Pi Pico}, KDP (2024).

\bibitem{key-5}
Thonny IDE, \url{https://thonny.org}

\bibitem{key-6}
RTL-SDR, \url{https://www.rtl-sdr.com}.

\bibitem{key-7}
GQRX, \url{https://www.gqrx.dk}

\bibitem{key-8}
GNU radio, \url{https://www.gnuradio.org/}

\bibitem{key-9}
Morse code, \url{https://en.wikipedia.org/wiki/Morse_code}

\bibitem{key-10}
Netcat, \url{https://en.wikipedia.org/wiki/Netcat}

\bibitem{key-11}
SoX, \url{https://en.wikipedia.org/wiki/SoX}

\bibitem{key-12}
multimon-ng, \url{https://github.com/EliasOenal/multimon-ng}

\bibitem{key-13}
buzzer\_music, \url{https://github.com/james1236/buzzer_music}.

\bibitem{key-14}
20th Century Fox theme, \url{https://www.moviesoundclips.net/sound.php?id=147}.

\bibitem{key-15}
FFmpeg, \url{https://ffmpeg.org}.

\bibitem{key-16}
MAX4466, \url{https://www.analog.com/media/en/technical-documentation/data-sheets/max4465-max4469.pdf}.



\end{thebibliography}
\end{document}